# Electron Scattering and Hybrid Phonons in Low Dimensional Laser Structures made with GaAs/Al$_x$Ga$_{1-x}$As.

V. N. Stavrou[1,2] and G.P. Veropoulos[1].

1. Division of Academic Studies, Hellenic Navy Petty Officers Academy, Skaramagkas, T.K. 12400, Greece
2. Department of Physics and Astronomy, University of Iowa, Iowa City, IA 52242, USA

**Abstract**

We theoretically and numerically present the hybrid phonon modes for the double heterostructure GaAs/Al$_x$Ga$_{1-x}$As and their interactions with electrons. More specifically, we have calculated the electron capture within a symmetric quantum well via the emission of hybrid phonons. Our investigation shows that the capture rates via the hybrid phonons are matched to the rates predicted by the dielectric continuum (DC) model and the concentration of aluminium which is an important parameter for controlling the electron capture process in light emitting diodes (LED).

1. **Introduction**

In the last two decades, low dimensional structures (LDS) have been studied. To be more specific, the electrical and optical properties of quantum wells (QWs), quantum wires (QWRs) and quantum dots (QDs) [1] have been a subject of thorough analysis. An application of LDS is the laser structures which have been in the center of the world wide research. The mechanism of carrier capture in LDS is of special importance for the operation of lasers based on QWs [2] and QDs [3]. Capture mechanism has been earlier studied [2, 4] by using the following approaches: a) well defined initial state [2] and b) considering an initial electron flux[4].

This paper is concerned with the study of the quantum carrier capture process, mediated by longitudinal optical phonons in heterostructure quantum wells based on III-V alloys semiconductors. In order to create a discrete energy spectrum for the electrons with energies greater than the barrier energy and in order to reduce electron-phonon scattering rates, we concentrate on the situation in which the heterostructure is confined between metallic barriers [5]. Calculations are then presented for the capture rates in the lowest electron subbands by the emission of hybrid phonons[6] in the case of the $GaAs/Al_xGa_{1-x}As$ heterostructure between two metal barriers. We have used the hybrid (HB) model to describe the optical phonons. The hybridons form an orthogonal set of modes and are characterised by the property of non-locality [6], which is identified as the source of the mechanical boundary conditions. The alloy system $Al_xGa_{1-x}As$ is mainly important for fabricating high speed electronic and optoelectronic devices because the lattice mismatch with GaAs is very small [7]. In order to investigate the effects of changing x and the well width, we have evaluated the electron capture rates for the double heterostructure $GaAs/Al_xGa_{1-x}As$ quantum well.

2. **Double hybrid and DC phonons.**

The hybrid model is a theoretical macroscopic model that has been successful in describing the polar optical modes in semiconductor low-dimensional structures [6]. The triple hybrid modes LO/TO/IP (where IP stands for the Interface Polariton type mode) satisfy both mechanical and electromagnetic boundary conditions, in contrast with the DC modes which satisfy only electromagnetic boundary conditions[8]. The TO modes may be neglected in this model since there is no electric field to contribute to the Fröhlich coupling mechanism [8] and, hence, TO modes can be discarded for the purpose of electron-phonon interactions. This consideration gives rise to the double hybrid model LO/IP [9].

The ionic displacement field **u** and the corresponding electric potential field $\Phi$ depend on the mechanical boundary conditions we choose. For the double heterostructure $GaAs/Al_xGa_{1-x}As$ it is suitable to use the following approximations:

(a) $\mathbf{u} \neq \mathbf{0}$ (well region) $\mathbf{u} \approx \mathbf{0}$ (barrier region)

(b) $\mathbf{u} \approx \mathbf{0}$ (well region) $\mathbf{u} \neq \mathbf{0}$ (barrier region)

Even when $\mathbf{u} \approx \mathbf{0}$, there is still an electric field associated with the IP type modes through the structure.

Before applying the above mentioned approximation for the optical phonons, it is worth mentioning the theoretical model for the dielectric function which will be used in our heterostructure. A theoretical approach for the description of the dielectric functions in QWs has been presented in Ref. [10]. The dielectric function for GaAs is given by:

$$\varepsilon_1(\omega) = \varepsilon_{\infty 1} \frac{\omega^2 - \omega_{L1}^2}{\omega^2 - \omega_{T1}^2} \tag{1}$$

while that for $Al_xGa_{1-x}As$ has a two-mode behaviour [7]:

$$\varepsilon_2(\omega) = \varepsilon_{\infty 2} \frac{\omega^2 - \omega_{L2,Ga}^2}{\omega^2 - \omega_{T2,Ga}^2} \frac{\omega^2 - \omega_{L2,Al}^2}{\omega^2 - \omega_{T2,Al}^2} \tag{2}$$

where the labels Ga and Al denote the GaAs-like and AlAs-like frequencies.

Considering case (a), the ionic displacement field in the quantum well region $|z| < L$ for the symmetric (S) modes can be written in terms of LO part (subscript L) and interface part (subscript I) as follows,

$$\mathbf{u}_{QW}^S = \left[ A_L^S \cos(k_{L1}z) + A_I^S \cosh(q_\| z), \quad 0, \quad A_L^S \frac{ik_{L1}}{q_\|} \sin(k_{L1}z) - A_I^S i \sinh(q_\| z) \right] \tag{3}$$

where we have omitted the common factor $e^{i(q_\| x - \omega t)}$, $\mathbf{q}_\|$ is the in-plane wavevector and the confinement wave vector $k_{L1}$ for the LO part is given by

$$\omega^2 = \omega_{L1}^2 - v_{L1}^2 (k_{L1}^2 + q_\|^2) \tag{4}$$

where $v_{L1}$ is the dispersive velocity of the LO mode in well material. In the same manner, we have found that the antisymmetric (A) modes can be written as

$$\mathbf{u}_{QW}^a = \left[ A_L^a \sin(k_{L1}z) + A_I^a \sinh(q_\| z), \quad 0, \quad -A_L^a \frac{ik_{L1}}{q_\|} \cos(k_{L1}z) - A_I^a i \cosh(q_\| z) \right] \tag{5}$$

where $A_L^{s,a}$ and $A_I^{s,a}$ are coefficients to be determined using the boundary conditions.

Applying the boundary conditions, the dispersion relation for the *symmetric modes* gets the form

$$\tanh(q_\| L) + s_1 \frac{k_{L1}}{q_\|} \tan(k_{L1}L) \left[ 1 + \frac{\varepsilon_1}{\varepsilon_2} \tanh(q_\| b) \tanh(q_\| L) \right] = 0 \tag{6}$$

where $b = D - L$. Similarly, for the *antisymmetric modes,* we obtain the dispersion relation

$$\coth(q_\| L) - s_1 \frac{k_{L1}}{q_\|} \cot((k_{L1}L)) \left[ 1 + \frac{\varepsilon_1}{\varepsilon_2} \tanh(q_\| b) \coth(q_\| L) \right] = 0 \tag{7}$$

The Hamiltonian for a hybrid mode is given by [1]

$$\hat{H} = \frac{M}{2V_c}\left[\int \hat{u}^2(\mathbf{r})d^3\mathbf{r} + \omega^2 \int \hat{u}^2(\mathbf{r})d^3\mathbf{r}\right] \qquad (8)$$

where $\hat{\mathbf{u}}(\mathbf{r})$ is the ionic displacement operator for the hybridons and $V_c$ is the crystal volume.

Considering case (b), the ionic displacement field in the barrier region $L < |z| < D$ for the symmetric (S) modes can be written as

$$\mathbf{u}_{BR}^s = \left[q_{\parallel}\left(A_1^s \cos(k_{L2}z) + A_2^s \sin(k_{L2}z)\right) + iq_{\parallel}\left(A_3^s e^{-q_{\parallel}|z|} + A_4^s e^{q_{\parallel}|z|}\right),\ 0,\right.$$
$$\left. k_{L2}\left((A_1^s \cos(k_{L2}z) + A_2^s \sin(k_{L2}z)\right) - q_{\parallel}\left(A_3^s e^{-q_{\parallel}|z|} + A_4^s e^{q_{\parallel}|z|}\right)\right] \qquad (9)$$

The antisymmetric mode displacement field, $\mathbf{u}_{BR}^a$, is defined in an analogous manner but with a minus sign when $z$ is negative.

The unknown coefficients can be determined by applying the boundary conditions at the interfaces and the normalization condition. The dispersion relations for the symmetric and antisymmetric modes can be written as

$$\left(\frac{s_2 k_{L2}}{q_{\parallel}}\right)^2 \sin(k_{L2}b)\left[\frac{\varepsilon_1}{\varepsilon_2}\sinh(q_{\parallel}b) + \cosh(q_{\parallel}b)\,\text{ct}(q_{\parallel}L)\right] -$$
$$\frac{s_2 k_{L2}}{q_{\parallel}}\left[2\frac{\varepsilon_1}{\varepsilon_2} - \cos(k_{L2}b)\left(2\frac{\varepsilon_1}{\varepsilon_2}\cosh(q_{\parallel}b) + \sinh(q_{\parallel}b)\,\text{ct}(q_{\parallel}L)\right)\right] - \qquad (10)$$
$$\frac{\varepsilon_1}{\varepsilon_2}\sinh(q_{\parallel}b)\sin(k_{L2}b) = 0,$$

where ct($x$) is coth($x$) for symmetric modes and tanh($x$) for antisymmetric modes.

The DC modes consists of two kind of modes: the confined and the interface modes [2]. The interaction Hamiltonian for the confined modes is given by

$$\hat{H}_C(\mathbf{r},z,t) = \begin{cases} \sum_l \int \left[C_{Conf2}(l,\mathbf{q}_{\parallel},z)e^{i(\mathbf{q}_{\parallel}\cdot\mathbf{r}-\omega t)}\hat{a}_{l,\mathbf{q}_{\parallel}} + HC\right]d^2\mathbf{q}_{\parallel} & -D \leq z \leq -L \\ \sum_k \int \left[C_{Conf1}(k,\mathbf{q}_{\parallel},z)e^{i(\mathbf{q}_{\parallel}\cdot\mathbf{r}-\omega t)}\hat{a}_{k,\mathbf{q}_{\parallel}} + HC\right]d^2\mathbf{q}_{\parallel} & |z| \leq L \\ \sum_m \int \left[C_{Conf2}(m,\mathbf{q}_{\parallel},z)e^{i(\mathbf{q}_{\parallel}\cdot\mathbf{r}-\omega t)}\hat{a}_{m,\mathbf{q}_{\parallel}} + HC\right]d^2\mathbf{q}_{\parallel} & L \leq z \leq D \end{cases} \qquad (11)$$

The confined modes have zero potential at the interfaces and have a frequency corresponding to the LO phonon of the material. In the same manner, the interaction Hamiltonian corresponding to interface modes could be given by:

$$\hat{H}_{IP}(\mathbf{r},z,t) = \sum_{j,k} \int \left[ C_{IP}^{j}(k,\mathbf{q}_{\parallel},z) e^{i(\mathbf{q}_{\parallel}\cdot\mathbf{r}-\omega t)} \hat{a}_{k,\mathbf{q}_{\parallel}} + HC \right] d^2\mathbf{q}_{\parallel} \quad (12)$$

where j denotes the symmetry and k specifies the IP mode. The coefficients in eq. (12) are determined by the normalization condition and the boundary conditions of the IP modes: the continuity at the inner interfaces and the vanishing at the outer interfaces.

## 3. Electron wavefunctions

We consider the orientation of the axes as shown in Fig.1 and solve the effective mass (finite $V_0$) Schrodinger equation for the electron wavefunction $\psi(\mathbf{r})$ allowed by the heterostructure with the energy eigenvalue E. Using the boundary conditions in which $\psi$ and $\frac{1}{m^*}\frac{\partial \psi}{\partial z}$ are continuous at the interfaces z=-L and z=L, vanishing the wavefunctions at the outer interfaces and taking into account the symmetry of the structure we have:

$$\psi_s(r_{\parallel},z) = A_s e^{ik_{\parallel}\cdot r_{\parallel}} \begin{cases} \sinh[k_2(D-L)]\cos(k_1 z), & |z| \leq L \\ \cos(k_1 L)\sinh[k_2(D-|z|)], & L \leq |z| \leq D \end{cases} \quad (13)$$

and

$$\psi_a(r_{\parallel},z) = A_a e^{ik_{\parallel}\cdot r_{\parallel}} \begin{cases} \sinh[k_2(D-L)]\sin(k_1 z), & L \leq |z| \leq D \\ \text{sgn}(z)\sin(k_1 L)\sinh[k_2(D-|z|)], & |z| \leq L \end{cases} \quad (14)$$

where the index s (a) corresponds to symmetric (antisymmetric) wavefunctions, $k_{\parallel}$ is a two dimensional wavevector along the interface planes and sgn (z) is the sign of z. Applying the boundary conditions, the symmetric and antisymmetric electron dispersion relations are given respectively by:

$$m_1^* k_2 \cos(k_1 L)\cosh[k_2(D-L)] - m_2^* k_1 \sin(k_1 L)\sinh[k_2(D-L)] = 0 \quad (15)$$

$$m_1^* k_2 \sin(k_1 L)\cosh[k_2(D-L)] + m_2^* k_1 \cos(k_1 L)\sinh[k_2(D-L)] = 0 \quad (16)$$

where $m_1^*$ and $m_2^*$ are the electron effective masses for GaAs and $Al_xGa_{1-x}As$ respectively.

## 4. Capture rates

The capture rates are defined as the sum of the transition rates for an electron from the bottom of the first subband above the well (initial state) to all possible bound states (final states) by emission of all possible polar optical modes [2]. The rates are calculated using the Fermi golden rule, assuming that only emission is possible. It is convenient to present capture rates in terms of the characteristic rate $\Gamma_0$ for bulk GaAs. We have used $\Gamma_0 = e^2 / 4\pi\varepsilon_0 \hbar \left(1/\varepsilon_{\infty 1} - 1/\varepsilon_{s1}\right)\sqrt{2m_1^* \omega_{L1}/\hbar}$ which is $8.7 \times 10^{12} \text{s}^{-1}$ using typical parameter values.

In figure 2(a), we present the capture rates dependence on the well width with fixed aluminium concentration $x$ using the Hybrid and DC phonons. When the electron states enter the well, the probability distribution for the electrons shifts from the barrier regions to the well and electron resonances appear because the overlap integrals in matrix elements[12] increase. The phonon resonances appear when the energy differences between the initial and the final state are equal to a phonon energy. At the phonon resonances, the initial electron wavefunction is located mainly in the barrier region [12].

For different aluminium concentration, the effective masses, energy gaps, LO and TO frequencies change. The dependence of rates on concentration $x$ with fixed well width is illustrated in figure 2(b). In the same manner as before, electron resonances appear when the states enter the well and phonon resonances are created when the energy differences are equal to phonon resonances. Figure 3 shows that small well width can produce large capture rates by changing the concentration $x$ and vice versa. Resonances can be controlled by the QWs characteristics –the well width $L$ and concentration $x$. This means that the alloy QW lasers can operate in a large range of wavelengths which depends on both $L$ and $x$.

5. **Conclusions**

We have calculated the electron capture rates for electrons localised at the bottom of the first subband above the well which make transitions into the quantum well (made with GaAs/Al$_x$Ga$_{1-x}$As) states by the emission of polar optical modes, as described by the hybrid and DC model. The evaluations have shown that the electron resonances occur at approximately the same well width and with comparable magnitudes for the above mentioned models. This has happened because the electron resonances are due to electron states and are not influenced by the model that describes the phonon emission. On the other hand, the phonon resonances, are influenced by the kind of polar optical phonons that are emitted. The resonances that appear in the case of the DC and hybrid phonons are reasonably well matched, as far as the magnitudes and the well width of the resonances are concerned.

6. **Acknowledgements**

The authors would like to thank Prof. M. Babiker, Prof. Brian Ridley for useful discussions and Lt. Commander of Hellenic Navy D. Filinis for his support. The author V.N.S. would like to acknowledge the financial support given by the Training Mobility of Researchers (TMR) and the University of Iowa under the grand No. 52570034.

## Figure captions

**Figure 1.** The profile illustrates a typical symmetric quantum well of width $2L$, inner barriers of width $(D-L)$ and infinite outer barriers at the points $-D$ and $D$.

**Figure 2.** The electron capture rates by the emission of Hybrid and DC [11] phonons for the heterostructure GaAs/Al$_x$Ga$_{1-x}$As for fixed $D$ ($D = 300$ Å) versus a) half the well width $L$ for $x=0.3$ and b) aluminium concentration $x$ for $L = 30$ Å.

**Figure 3.** The electron capture rates by the emission of Hybrid phonons for the heterostructure GaAs/Al$_x$Ga$_{1-x}$As for fixed $D$ ($D = 300$ Å) versus half the well width $L$ and aluminium concentration $x$.

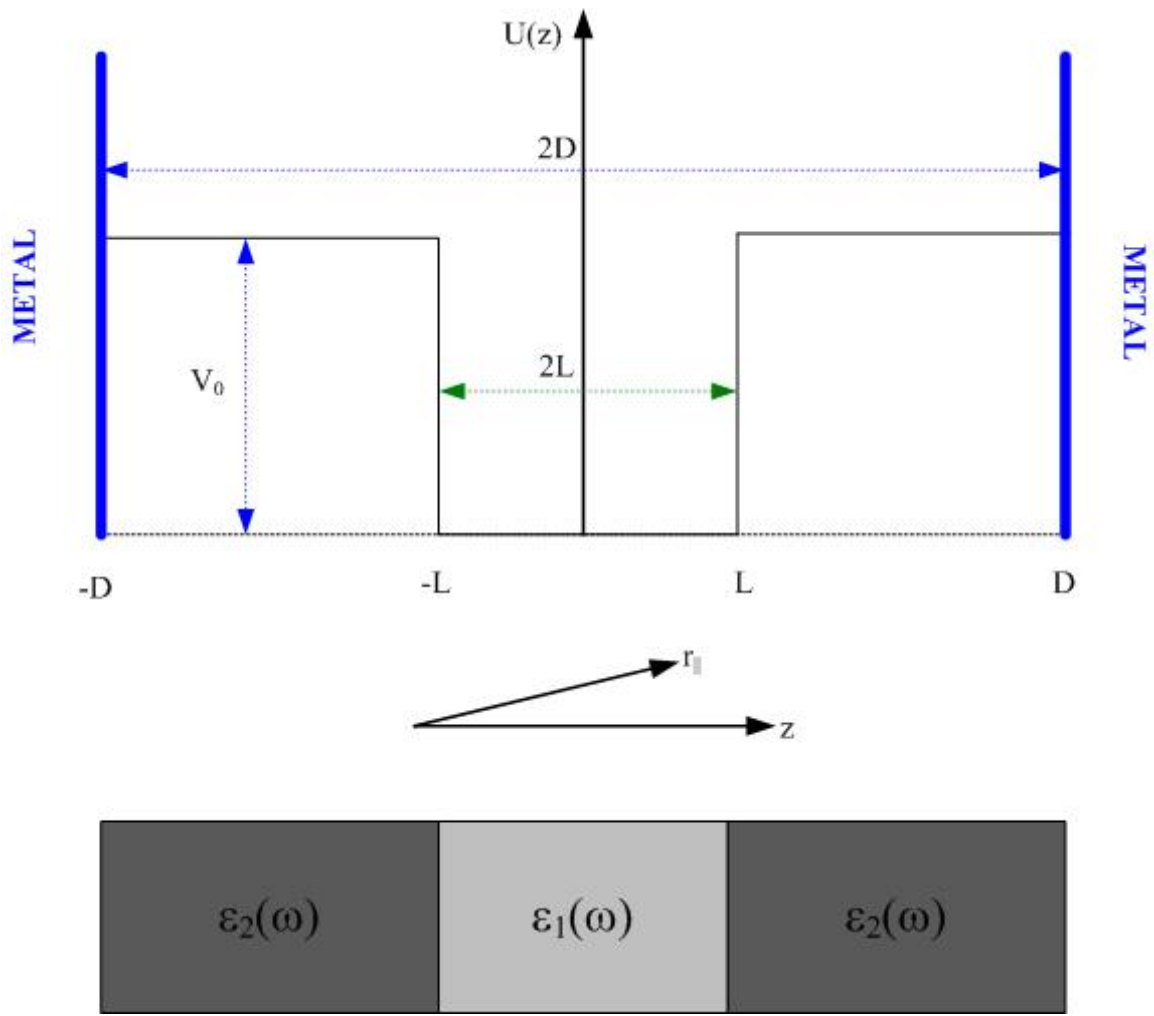

**Figure 1.** V. N. Stavrou & G.P. Veropoulos

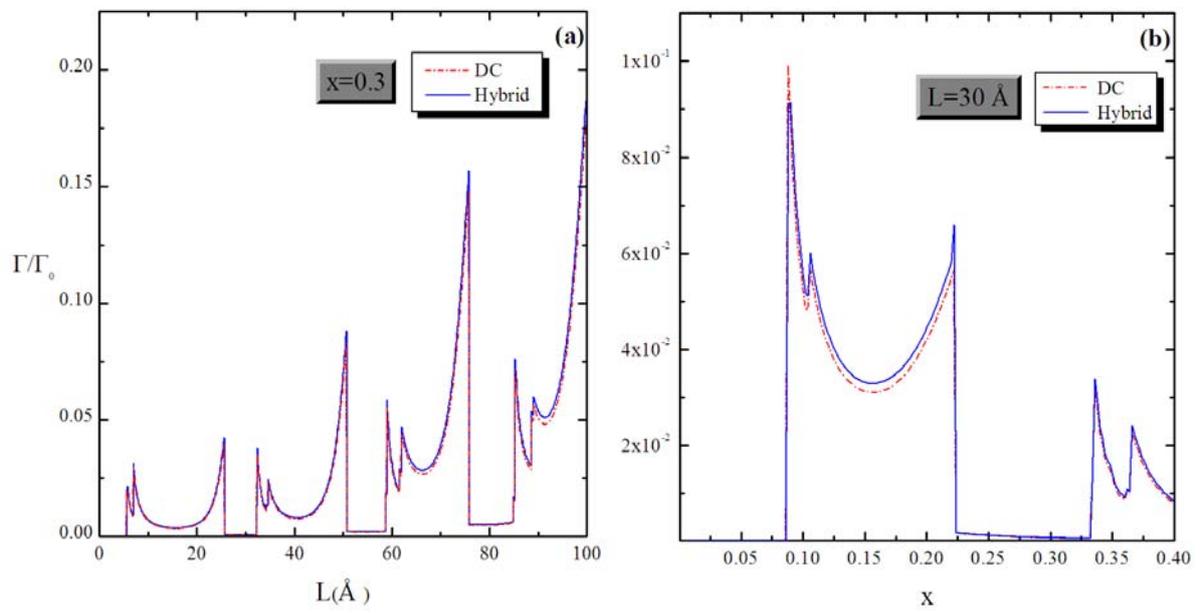

**Figure 2.**     V. N. Stavrou & G.P. Veropoulos

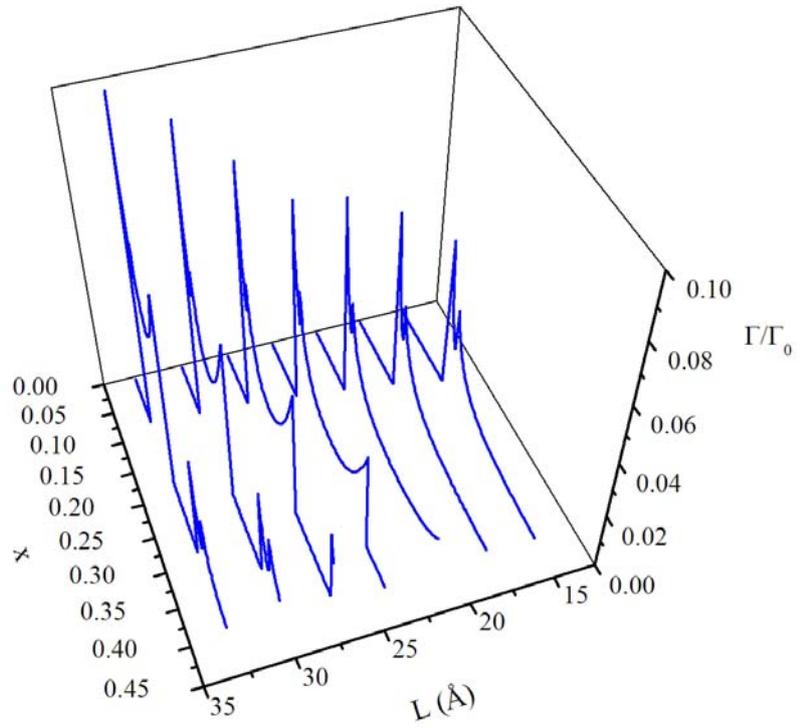

**Figure 3.**     V. N. Stavrou & G.P. Veropoulos